\documentclass[%
aps,preprint,prc,amsmath,amssymb,nofootinbib,longbibliography,superscriptaddress%
]{revtex4-1}

\usepackage[utf8]{inputenc}
\usepackage[usenames]{color}
\usepackage[dvipsnames]{xcolor}
\usepackage{graphicx}
\usepackage{amsmath}
\usepackage{amsfonts}
\usepackage{amssymb}
\usepackage{mathrsfs}
\usepackage{bm}
\usepackage{verbatim}
\usepackage{cancel}
\usepackage{comment}
\usepackage[normalem]{ulem}
\usepackage{float}
\usepackage{textcomp}
\usepackage{booktabs}
\usepackage{braket} 

 \usepackage{hyperref}
 \hypersetup{
    pdfnewwindow=true,      
   colorlinks=true,        
   linkcolor=ForestGreen,    
   citecolor=PineGreen,    
  filecolor=PineGreen,    
  urlcolor=PineGreen      
 }

\usepackage{multirow,makecell}

\newcommand{\mhi}{M_{\text{hi}}}

\newcommand{\chn}[3]{{{}^{#1}\!{#2}_{#3}}}
\newcommand{\cs}[2]{\chn{#1}{S}{#2}}
\newcommand{\cp}[2]{\chn{#1}{P}{#2}}
\newcommand{\cd}[2]{\chn{#1}{D}{#2}}
\newcommand{\cf}[2]{\chn{#1}{F}{#2}}

\newcommand{\cpf}{{\cp{3}{2}-\cf{3}{2}}}

\newcommand{\NNLO}{N$^2$LO}
\newcommand{\NNNLO}{N$^3$LO}
\newcommand{\NNNNLO}{N$^4$LO}

\newcommand{\Mpisq}{M^2_{\pi}}

\graphicspath{{./figs/}}

\begin{document}


\title{Enhancement of deltaful two-pion exchange nuclear forces}

\author{Haiming Chen}
\affiliation{College of Physics, Sichuan University, Chengdu, Sichuan 610065, China}

\author{Rui Peng}
\affiliation{School of Physics, Peking University, Beijing 100871, China}

\author{Songlin Lyu}
\email{songlin.lyu@na.infn.it}
\affiliation{College of Physics, Sichuan University, Chengdu, Sichuan 610065, China}

\author{Bingwei Long}
\email{bingwei@scu.edu.cn}
\affiliation{College of Physics, Sichuan University, Chengdu, Sichuan 610065, China}
\affiliation{Southern Center for Nuclear-Science Theory (SCNT), Institute of Modern Physics, Chinese Academy of Sciences, Huizhou 516000, Guangdong Province, China}

\date{April 11, 2024}

\begin{abstract}
The role of the delta isobar degrees of freedom in nucleon-nucleon scattering is revisited. We attempt to understand why the dimensionally regularized two-pion exchanges with the explicit delta isobar is much stronger than the ones with spectral function regularization. When the cutoff value of spectral function regularization is varied, the isoscalar central component exhibits a rather large cutoff variation. This reveals a surprisingly large numerical factor of the deltaful two-pion exchange potentials. The power counting is adjusted accordingly and we discuss the results and how to improve upon this finding.
\end{abstract}

\maketitle


\section{Introduction\label{sec:intro}}

The delta-isobar resonance $\Delta(1232)$ is the closest baryon to the nucleon, marked by the nucleon-delta mass splitting $\Delta \simeq 293$ MeV, thus it has been long thought to contribute significantly to two-nucleon~\cite{Ordonez:1995rz, Kaiser_1998, Krebs:2007rh, Schiavilla2015, Schiavilla2016} and three-nucleon interactions~\cite{Fujita:1957zz,Coon1993, vanKolck:1994yi, Epelbaum:2007sq, Kres2018:delta}, in addition to other scenarios of improving chiral nuclear forces~\cite{Epelbaum2011:rel, Long2012:RcnfT, Long2012:Srs, Long2013:1s0, Valderrama2017, Mario2018:1s0, Ren_2018, Long2022:separable, Papenbrock2022:TPEasLO, Ren:2022glg, Geng2022:RNNI, Lu:2022yxb}. Besides derivation of chiral nuclear forces, there are efforts to investigate the role of the delta isobar in nuclear structure by accounting for these so-called deltaful chiral forces in ab initio calculations~\cite{Jiang:2020the, Hu:2021trw, Bonaiti2022}. We are motivated in part by a recent analysis of a delta-less, renormalization-group (RG) invariant chiral forces~\cite{Thim:2024yks}, where scattering data are described by such a power counting only up to the on-shell centre of mass (CM) momentum $k \simeq 200$ MeV. Although disappointing, the conclusion of the perturbative analysis is consistent with expectation of a deltaless power counting: it converges in a kinematic window limited by $k \lesssim \Delta$. This is evident from the factor $1/\Delta$ in deltaful two-pion-exchange (TPE) Eq.~\eqref{eqn:TPEVCbc}.

In this paper we report the findings in our first investigation on the role of the delta isobar in nucleon-nucleon scattering. In particular, we are interested to know why the deltaful TPE potentials differ so significantly between two regularization schemes:  dimensional regularization (DR) and spectral function regularization (SFR). If the DR is used~\cite{Kaiser_1998}, the TPE diagrams shown in Fig.~\ref{fig:OneDeltaBox} do not yield any explicit regularization-scale dependence, once the divergences are subtracted by constant or momentum-squared counterterms. It has become known in the literature that the DR version of the deltaful TPE forces appears to be overly attractive in comparison with the phase shifts from the partial-wave analyses (PWA) \cite{nn.online, Stoks:1993tb}. The SFR scheme~\cite{Epelbaum_2004, Krebs:2007rh}, however, does have an explicit momentum cutoff $\tilde{\Lambda}$. If $\tilde{\Lambda}$ is taken to the limit $\tilde{\Lambda} \to \infty$, the DR expressions will be reproduced, at least for these deltaful TPE nucleon-nucleon ($NN$) diagrams. $\tilde{\Lambda} = 700$ MeV were advocated because it leads to better agreement with the PWA. This setup was popular in the various applications of deltaful chiral forces, including $NN$ scattering~\cite{Epelbaum2008,Valderrama2009,Machleidt2021,Mishra2022}, finite nuclei~\cite{Jiang:2020the} and infinite nuclear matter~\cite{Schwartz2018}. Among the results, the Goteborg-Oak Ridge (GO) group reported promising results on finite nuclei and nuclear matter with deltaful chiral potential~\cite{Jiang:2020the}. However, the NN scattering phase shift was found to be poorly reproduced, as highlighted in Ref.~\cite{Machleidt2021}.

Adopting a particular choice of the cutoff value is equivalent to correlating the series of $NN$ contact forces, or counterterms. Since $\mathcal{O}(Q^2)$ counterterms with adjustable low-energy constants (LECs) are already included at the same level with TPEs in power counting, the said choice of $\tilde{\Lambda}$ implies that a combination of $\mathcal{O}(Q^4)$ counterterms, thought to be higher-order in naive dimensional analysis (NDA), are included implicitly and their values are related to $\tilde{\Lambda} = 700$ MeV. If they are indeed important for the deltaful TPE to agree with $NN$ scattering data, there must be a rationale to power count them differently than in NDA. Therefore, understanding $Q^4$ counterterms subsumed in the SFR at finite $\tilde{\Lambda}$ is the lead at the beginning of our investigation. 

On the other hand, the TPE diagrams are computed in complete perturbation theory, and cutoff dependence of Feynman diagrams is known to follow NDA. As we will see, it is actually the numerical factor of certain diagrams that are less suppressed than estimated by $1/(4\pi)^{\nu}$ where $\nu$ is the chiral index defined by Weinberg~~\cite{Weinberg:1990rz, Weinberg:1991um}. In addition, appearance of $\Delta$ in the denominator also helps because $Q \sim \Delta$ is the kinematic window we are interested in, where $Q$ denotes generically the size of momenta involved in the processes. We will incorporate this enhancement into power counting and re-examine the role of the delta isobar by studying the peripheral partial-wave phase shifts. 

The paper is organized as follows. In Sec.~\ref{sec:theory} we identify the most dominant deltaful TPE force by studying the residual cutoff dependence. After adjusting the power counting, we compare with the empirical $NN$ phase shifts in Sec.~\ref{sec:results}. Finally, discussions and a summary are offered in Sec.~\ref{sec:disc}.  

\section{Enhancement of deltaful TPE potentials~\label{sec:theory}}

Following the conventions of Ref.~\cite{Kaiser_1998}, we break down the $NN$ forces to a combination of spin and isospin operators:
\begin{align}
\label{VNN}
V(\vec{p}\,', \vec{p}) &=V_C + \bm{\tau}_1 \bm{\cdot} \bm{\tau}_2 W_C + \left(V_S + \bm{\tau}_1\cdot\bm{\tau}_2W_S\right)\Vec{\sigma}_1\cdot\Vec{\sigma}_2 \nonumber \\
    & \quad + \left(V_T + \bm{\tau}_1\cdot\bm{\tau}_2W_T\right)\Vec{\sigma}_1\cdot\Vec{q}\, \Vec{\sigma}_2\cdot\Vec{q} \, ,
\end{align}
where $\vec{p}$ ($\vec{p}\,'$) is the initial (final) relative momentum and $\vec{q} \equiv \vec{p}\,' - \vec{p}$ is the momentum transfer. $V_i$ and $W_i$ are scalar functions of $\vec{p}$ and $\vec{p}\,'$. For instance, the one-pion-exchange (OPE) potential is given by
\begin{equation}
    V_{\mathrm{OPE}}(\Vec{p}\,',\Vec{p}) = W_T^{(\pi)}(q) \bm{\tau}_1\cdot\bm{\tau}_2 \Vec{\sigma}_1\cdot\Vec{q}\, \Vec{\sigma}_2\cdot\Vec{q} \, ,
\end{equation}
and
\begin{equation}
W_T^{(\pi)}(q) \equiv 
    -\frac{g^2_A}{4f^2_\pi}\frac{1}{q^2+\Mpisq}\, , \label{eqn:OPEWT}
\end{equation}
where the pion decay constant $f_\pi = 92.4$ MeV, nucleon axial-vector coupling $g_A = 1.29$, pion mass $M_\pi = 138$MeV. In both $S$ waves and at least part of $P$ waves~\cite{Bira2005:reOPE, Wu:2018lai, Kaplan:2019znu}, it has been established that OPE is nonperturbative for $k \simeq \Delta$. Therefore, OPE is counted as leading order (LO) because it must be iterated nonperturbatively. In this paper, however, we will study peripheral partial waves where OPE is perturbative, therefore, it is considered NLO in those channels:
\begin{equation}
    V_\text{NLO} = V_\text{OPE} \, .
\end{equation}

\begin{figure}[htb]
    \centering
    \includegraphics[scale=0.8]{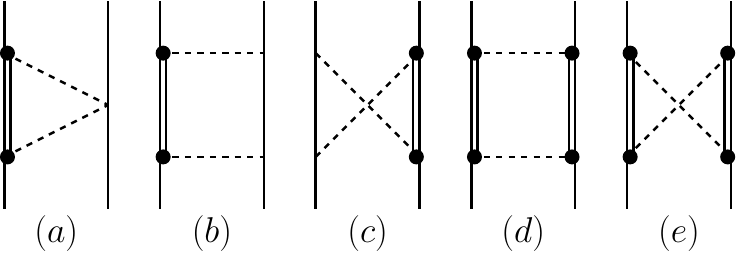}
    \caption{
    The leading TPE potentials with $\Delta$ intermediate states. The dashed, solid, and double-solid lines correspond to the pion, the nucleon, and the $\Delta$ resonance, respectively.}
    \label{fig:OneDeltaBox}
\end{figure}

The leading deltaful TPE diagrams are those with chiral index $\nu = 0$, as shown in Fig.~\ref{fig:OneDeltaBox}. To avoid double counting contributions that are already accounted for at LO through resummation of OPE, one must subtract contributions from purely two-nucleon intermediate states. Some of the physical constants needed are the $N \Delta$ axial-coupling $h_A = 1.40$, the nucleon mass $M_N = 939$ MeV, and the delta-nucleon mass splitting $\Delta = M_{\Delta} - M_N = 293.7$ MeV. The analytic expressions have been worked out in Ref.~\cite{Kaiser_1998} with DR and in Ref.~\cite{Krebs:2007rh} with SFR. The SFR scheme is a dispersion integral, with the regularized potentials being obtained through an integral over spectral functions
\begin{equation}
    V(q) = \frac{2}{\pi} \int_{2m_\pi}^\infty \mathrm{d} \mu \, \mu \frac{\rho(\mu; \tilde{\Lambda})}{\mu^2 + q^2}
\end{equation}
where the spectral function $\rho(\mu, \tilde{\Lambda})$ is derived by completing the TPE Feynman diagrams with a sharp momentum cutoff, see more detail in Ref.~\cite{Epelbaum_2004}.

We categorize these potentials in three groups, as done in both papers. In the first group, the potential is composed of the triangle diagram Fig.~\ref{fig:OneDeltaBox}(a) and its variants by permutation of the vertexes:
\begin{align}
    &W_{C}^{(a)}
=-\frac{h^2_A}{216\pi^2 f^4_{\pi}}\left[(5q^2+8M^2_\pi-12\Delta^2)L^{\tilde{\Lambda}}(q)+12\Delta^2(q^2+2M^2_\pi-2\Delta^2)D^{\tilde{\Lambda}}(q)\right] \, .
\label{eqn:TPEa}
\end{align}
The second group are the box diagrams with a single-$\Delta$ excitation Fig.~\ref{fig:OneDeltaBox}~(b) and (c):

\begin{align}
&V_{C}^{(b + c)}
=-\frac{g^2_Ah^2_A}{12\pi\Delta f^4_{\pi}}\left(q^2+2M^2_\pi\right)^2A^{\tilde{\Lambda}}(q) \, , \label{eqn:TPEVCbc} \\
&W_{C}^{(b+c)}
=-\frac{g^2_Ah^2_A}{216\pi^2 f^4_{\pi}}\left[(12\Delta^2-20M^2_\pi-11q^2)L^{\tilde{\Lambda}}(q)+6(q^2+2M^2_\pi-2\Delta^2)^2D^{\tilde{\Lambda}}(q)\right]\, ,\\
&V_{T}^{(b+c)}=-\frac{1}{q^2}V_{S}^{(b+c)}=-\frac{g^2_Ah^2_A}{48\pi^2 f^4_{\pi}}\left[-2L^{\tilde{\Lambda}}(q)+(q^2+4M^2_\pi-4\Delta^2)D^{\tilde{\Lambda}}(q)\right]\, ,\\
&W_{T}^{(b+c)}=-\frac{1}{q^2}W_{S}^{(b+c)}=-\frac{g^2_Ah^2_A}{144\pi \Delta f^4_{\pi}}\left(q^2+4M^2_\pi\right)A^{\tilde{\Lambda}}(q)\, .
\label{eqn:TPEbc}
\end{align}

The double-$\Delta$ excitation diagrams Fig.~\ref{fig:OneDeltaBox}~(d) and (e) make up the third group:

\begin{align}
&V_{C}^{(d + e)}
=-\frac{h^4_A}{27\pi^2 f^4_{\pi}}\Bigg\{ -4\Delta^2L^{\tilde{\Lambda}}(q)+(q^2+2M^2_\pi-2\Delta^2) \nonumber \\
&\qquad \qquad \qquad \qquad \qquad  \qquad \qquad \times \left[H^{\tilde{\Lambda}}(q)+(q^2+2M^2_\pi+6\Delta^2)D^{\tilde{\Lambda}}(q)\right]\Bigg\} \, ,\\
&W_{C}^{(d+e)}
=-\frac{h^4_A}{486\pi^2 f^4_{\pi}}\Bigg\{(11q^2+20M^2_\pi-24\Delta^2)L^{\tilde{\Lambda}}(q) \nonumber \\
&\qquad \qquad \qquad \qquad +3(q^2+2M^2_\pi-2\Delta^2)\left[H^{\tilde{\Lambda}}(q) + (-q^2-2M^2_\pi+10\Delta^2)D^{\tilde{\Lambda}}(q)\right]\Bigg\}\, ,\\
&V_{T}^{(d+e)}=-\frac{1}{q^2}V_{S}^{(d+e)}=-\frac{h^4_A}{216\pi^2 f^4_{\pi}}\left[6L^{\tilde{\Lambda}}(q)+(-q^2-4M^2_\pi+12\Delta^2)D^{\tilde{\Lambda}}(q)\right]\, ,\\
&W_{T}^{(d+e)}=-\frac{1}{q^2}W_{S}^{(d+e)}=-\frac{h^4_A}{1296\pi^2 f^4_{\pi}}\left[2L^{\tilde{\Lambda}}(q)+(q^2+4M^2_\pi+4\Delta^2)D^{\tilde{\Lambda}}(q)\right]\, .
\label{eqn:TPEde}
\end{align}
Here the momentum cutoff $\tilde{\Lambda}$ is related to the SFR scheme. Various functions of $q$ and $\tilde{\Lambda}$ used above are defined as follows:
\begin{align}
&L^{\tilde{\Lambda}}(q) =  \frac{\sqrt{q^2+4M^2_\pi}}{2q} \nonumber \\
&\qquad \times \ln{\frac{\tilde{\Lambda}^2\left(q^2+4M^2_\pi\right)+q^2(\tilde{\Lambda}^2-4M^2_\pi)+2\tilde{\Lambda} q\sqrt{\left(q^2+4M^2_\pi\right)(\tilde{\Lambda}^2-4M^2_\pi)}}{4M^2_\pi(\tilde{\Lambda}^2+q^2)}}\, ,\\
&A^{\tilde{\Lambda}}(q) =  \frac{1}{2q}\arctan{\frac{q(\tilde{\Lambda}-2M_\pi)}{q^2+2\tilde{\Lambda} M_\pi}}\, ,\\
&D^{\tilde{\Lambda}}(q) =  \frac{1}{\Delta} \int_{2M_\pi}^{\tilde{\Lambda}}\frac{\mathrm{d}u}{u^2+q^2} \arctan{\frac{\sqrt{u^2-4M^2_\pi}}{2\Delta}}\, ,\\
&H^{\tilde{\Lambda}}(q) = \frac{2q^2+4M^2_\pi-4\Delta^2}{q^2+4M^2_\pi-4\Delta^2}\left[L\left(q\right)-L\left(2\sqrt{\Delta^2-M^2_\pi}\right)\right]\, .
\label{LADH}
\end{align}

The choice of $\tilde{\Lambda}$ is arbitrary as long as it is higher than the breakdown scale of ChEFT $\mhi$. The dependence of the $NN$ scattering amplitude on $\tilde{\Lambda}$ should be no more important than higher-order effects, a requirement equivalent to renormalization-group (RG) invariance. The $\tilde{\Lambda}$ dependence of the deltaful TPE potential listed in Eqs.~\eqref{eqn:TPEa}-\eqref{eqn:TPEde} is ``residual'' in the sense that it vanishes at the limit $\tilde{\Lambda} \to \infty$. In fact, one arrives at the DR expressions given in Ref.~\cite{Kaiser_1998} for $\tilde{\Lambda} \to \infty$. 

Using the analytical forms of Eqs.~\eqref{eqn:TPEa}-\eqref{eqn:TPEde} and $m_\pi \sim q \sim \Delta \ll \tilde{\Lambda}$, one can expand the residual cutoff dependence in powers of $1/\tilde{\Lambda}$. These $\tilde{\Lambda}^{-n}$ terms all appear as polynomials in $q^2$, that is, as $NN$ contact forces. We focus on the $q^4$ terms, because, as mentioned in Sec.~\ref{sec:intro}, polynomials up to $q^2$ are already included at the same order as TPEs. Therefore, the $q^4$ counterterms are the first in the line to potentially upset NDA. We find that the most dominant piece of the residual cutoff dependence is contributed by $V_C^{(b+c)}$ and $V_C^{(d+e)}$: 
\begin{equation}
    V_C^{(q^4)} = \frac{g^2_A h^2_A}{ 24\pi f^4_{\pi} \Delta \tilde{\Lambda}} \left(1 + \frac{4}{9} \frac{h_A^2}{g_A^2} \right) q^4 \, .
    \label{eqn:VCq4}
\end{equation}
By comparison, the second largest term, $W_C^{(q^4)}$, is smaller by a factor of approximately $15\%$. With the usual scaling for ChEFT with the explicit delta degrees of freedom~\cite{Pascalutsa2003_05, Pascalutsa2003_11, Pandharipande2005, LONG2010:delta}, 
\begin{align}
    4 \pi f_\pi \sim \tilde{\Lambda} \sim \mhi \, , q \sim f_\pi \sim \Delta \sim Q \, ,
\end{align}
$V_C^{(q^4)}$ is counted as
\begin{equation}
V_C^{(q^4)} \sim 
\frac{1}{\mhi^2} \, .
\end{equation}
We notice that $V_{C}^{(q^4)}$ is smaller than OPE by $\mathcal{O}(Q^2/\mhi^2)$. Its size becomes more considerable if one accounts for the numerical fact $q \sim 3 f_\pi$ because we are interested in the kinematic region $q \sim \Delta$. At any rate, the unusual significance of this residual term suggests that the long-range part, i.e., the terms that survives at $\tilde{\Lambda} \to \infty$, of the isoscalar central force
\begin{equation}
    V_{C-\Delta}^{\text{TPE}} \equiv V_C^{(b+c)} + V_C^{(d+e)}
\end{equation}
is even larger, and it is smaller than OPE by only one order. However, $V_{C-\Delta}^{\text{TPE}}$ is counted in NDA two-order smaller than OPE. 
It follows that for peripheral waves interested in the paper, the central part of the deltaful TPE must be considered {\NNLO}, compared with its NDA counting of {\NNNLO}. Development of ChEFT has seen before enhancement of certain diagrams due to unexpectedly large numerical factors. For instance, in Refs.~\cite{Liebig:2010ki, Baru:2012iv}, the seagull diagrams of deltaless TPEs with $\nu = 1$ vertexes appear to be enhanced too. Similar to the numerical enhancement of the seagull diagrams, the denominator in Eq.~\eqref{eqn:VCq4} is proportional to $\pi$, as opposed to $\pi^2$ estimated by the standard ChEFT counting. 

It is not totally clear to us where the origin of this enhancement lies. Instead, we proceed to examine its consequence in peripheral partial waves of $NN$ scattering. It is instructive to review the perturbative counting for peripheral waves without complication due to $\Delta$, i.e., power counting with the deltaless TPEs. As mentioned earlier, since none of the forces are treated nonperturbatively, the LO potential is considered vanishing and OPE appears at NLO. As argued in Ref.~\cite{Wu:2018lai}, the {\NNLO} potential includes the $\mathcal{O}(Q^2)$ counterterms and all the deltaless TPEs appear at {\NNNLO}. The deltaless counting scheme is tabulated as follows:
\begin{equation}
\begin{aligned}
\mathrm{LO}:\ V_{\mathrm{LO}}       &= 0\, ,\\
\mathrm{NLO}:\ V_{\mathrm{NLO}}     &= V_{\mathrm{OPE}}\, ,\\
\mathrm{N^2LO}:\ V_{\mathrm{N^2LO}} &= V^{(2)}_{\mathrm{ct}}\, ,\\
\mathrm{N^3LO}:\ V_{\mathrm{N^3LO}} &= V^{\text{TPE} \cancel{\Delta}} \, ,
\label{PW:deltaless}    
\end{aligned}
\end{equation}
where $V^{\text{TPE} \cancel{\Delta}}$ refers to the TPE potentials with $\nu = 0$ vertexes without the $\Delta$ field and $V^{(2)}_{\mathrm{ct}}$ represent the $\mathcal{O}(Q^2)$ counterterms acting in the $P$ waves
\begin{equation}
    \bra{\mathrm{chn},p'}V^{(2)}_{\mathrm{ct}}\ket{\mathrm{chn},p} = C_{\mathrm{chn}}pp' \, ,
\end{equation}
with $\mathrm{chn} = {}^1P_1, {}^3P_1, {}^3P_2$. But we refrain from applying to the $P$ wave because it is not certain whether OPE is still perturbative at the relative momentum $k \simeq \Delta$.

Now with the enhancement discovered above, the power counting of the deltaful TPEs needs to be adjusted. The central force of the deltaful TPEs $V_{C-\Delta}^{\text{TPE}}$ must be added to {\NNLO}. We then have the rest of the TPEs and $V_C^{(q^4)}$ at {\NNNLO}:
\begin{equation}
\begin{aligned}
\mathrm{LO}:\ V_{\mathrm{LO}}       &= 0\, ,\\
\mathrm{NLO}:\ V_{\mathrm{NLO}}     &= V_{\mathrm{OPE}}\, ,\\
\mathrm{N^2LO}:\ V_{\mathrm{N^2LO}} &= 
V_{C-\Delta}^{\text{TPE}}\, ,\\
\mathrm{N^3LO}:\ V_{\mathrm{N^3LO}} &= 
V_C^{(q^4)} + \, \text{Rest of TPEs} \, .
\label{PW:deltaVCq4}    
\end{aligned}
\end{equation} 
For definiteness, the TPE potentials quoted above all use the DR version, or $\tilde{\Lambda} \to \infty$ in its SFR formulation. When the central counterterm $V_C^{(q^4)}$ is projected onto the $D$ waves, the numerical prefactor is independent of the spin $s$ and total angular momentum $j$:
\begin{align}  
\langle D\, s\, j, p' | V_C^{(q^4)} | D\, s\, j, p \rangle &= \frac{8}{15} p^2{p^{\prime}}^2 D_{C-\Delta} \, , 
\end{align}  
where the value of $D_{C-\Delta}$ is parametrized by $\tilde{\Lambda}$:
\begin{equation}
    D_{C-\Delta} = \frac{g^2_A h^2_A}{ 24\pi f^4_{\pi} \Delta \tilde{\Lambda}} \left(1 + \frac{4}{9} \frac{h_A^2}{g_A^2} \right) \, .
    \label{eqn:DefDCq4}
\end{equation}

Computation of the $NN$ $T$ matrix according to a perturbative counting is explained in detail in Refs.~\cite{Wu:2018lai, Kaplan:2019znu}. In a nutshell, $n$ insertions of order-$\nu$ potentials contributes to the $T$ matrix at order-$n \nu$. Therefore, 
\begin{align*}
    T_\text{LO} &= 0 \, , \\
    T_\text{{NLO}} &= V_\text{NLO} \, , \\
    T_\text{{\NNLO}} &= V_\text{\NNLO} + V_\text{NLO} G_0 V_\text{NLO} \, , \\
    T_\text{{\NNNLO}} &= V_\text{\NNNLO} + V_\text{NLO} G_0 V_\text{NLO} G_0 V_\text{NLO}\nonumber \\
    &\quad + V_\text{\NNLO} G_0 V_\text{NLO} + V_\text{NLO} G_0  V_\text{\NNLO} \, ,
\end{align*}
where $G_0$ is the free two-nucleon propagator. Iterations of the sort $V G_0 V$ implies summing over $NN$ intermediate states, which is done in the partial-wave projected form:
\begin{equation}
    \int \frac{\mathrm{d} l l^2}{2\pi^2} V(p', l) \frac{m_N}{k^2 - l^2 + \mathrm{i}\epsilon} V(l, p)\, .
\end{equation}
They are regularized by the separable Gaussian function:
\begin{equation}
    V(p\,', p; \Lambda) \equiv \mathrm{e}^{-\left(\frac{p'}{\Lambda}\right)^4}  V(p\,', p\,) \mathrm{e}^{-\left(\frac{p}{\Lambda}\right)^4}  \, .
\end{equation}
The cutoff $\Lambda$, which regularizes $NN$ intermediate states in the distorted-wave expansion, is to be distinguished from $\tilde{\Lambda}$ used in the SFR, which regularized the diagrams of Fig.~\ref{fig:OneDeltaBox}. The dependence on $\Lambda$ turns out to be much less sensitive than that on $\tilde{\Lambda}$, therefore we fix its value $\Lambda = 800$ MeV throughout the paper.

\section{Results\label{sec:results}}

\begin{figure}[htb]
    \centering
    \includegraphics[scale=0.49]{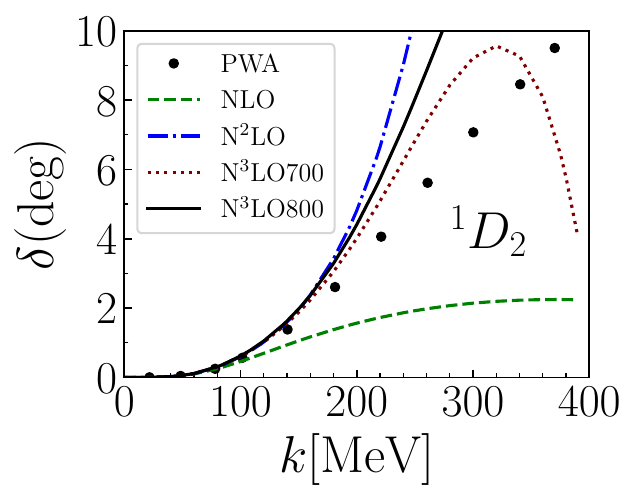}
    \includegraphics[scale=0.49]{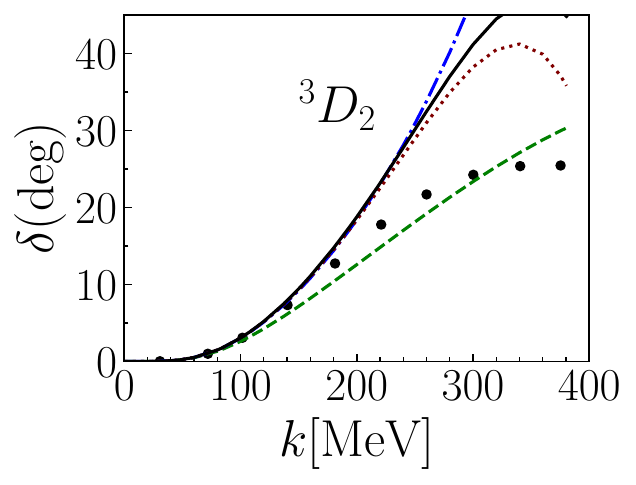}
    \includegraphics[scale=0.49]{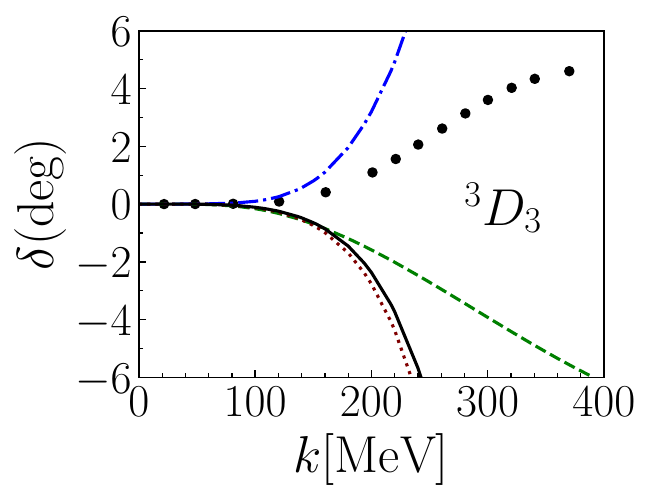}
    \caption{$NN$ phase shifts as functions of center-of-mass (CM) momentum in various $D$ waves with the deltaful TPEs. ``PWA'' refers to the empirical phase shifts from Ref.~\cite{nn.online}. The solid and dotted curves are plotted with different values of $D_{C-\Delta}$. See the text for more explanation.
    }
    \label{fig:d_newC4}
\end{figure}

\begin{figure}[htb]
    \centering
    \includegraphics[scale=0.49]{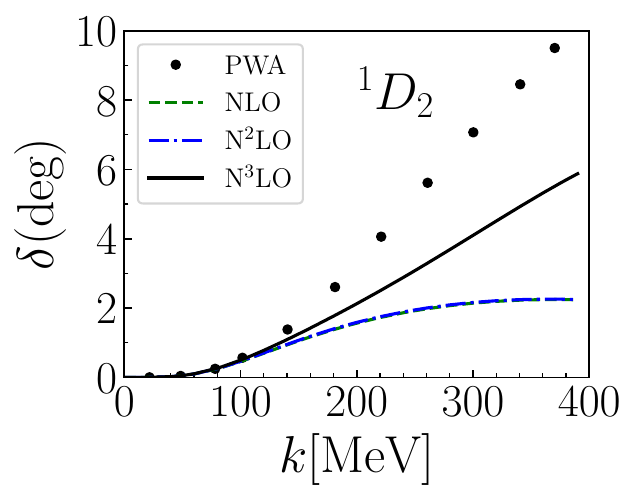}
    \includegraphics[scale=0.49]{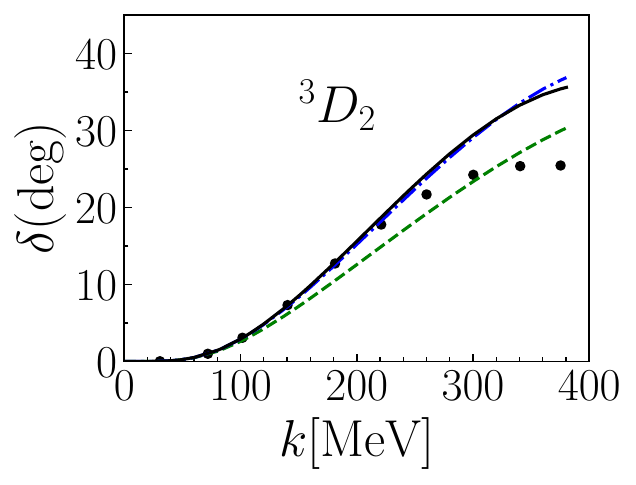}
    \includegraphics[scale=0.49]{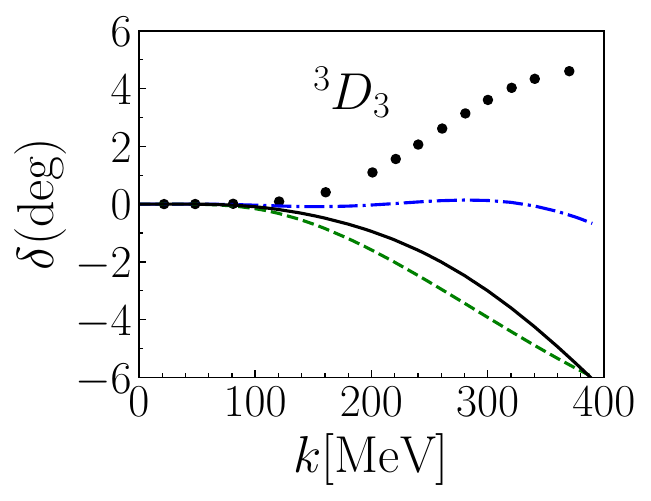}
    \caption{$NN$ phase shifts of the $D$ waves up to {\NNNLO} with the \emph{deltaless} TPEs.}
    \label{fig:d_deltaless}
\end{figure}

We first look at the $D$ waves. The perturbative counting is not applied to $\cd{3}{1}$ because $\cd{3}{1}$ is coupled to $\cs{3}{1}$ and OPE contributes strongly to this coupled channel. The phase shifts produced by the deltaful TPEs are shown in Fig.~\ref{fig:d_newC4}. We use $\tilde{\Lambda} = 700$ and $800$ MeV in Eq.~\eqref{eqn:DefDCq4} to represent difference choice of $D_{C-\Delta}$. The phase shifts generated by the deltaless TPE potentials according to power counting \eqref{PW:deltaless} are plotted in Fig.~\ref{fig:d_deltaless} for comparison. As $V_{\mathrm{ct}}^{(2)}$ vanishes in $D$ and $F$ waves, there are no free parameters. We observe that $V^{TPE}_{C-\Delta}$ indeed takes up much of the strength of the deltaful TPEs. This is evident in $\cd{1}{2}$ and $\cd{3}{2}$ where {\NNLO} appears to be close to {\NNNLO}. However, this is not quite the case in $\cd{3}{3}$, where {\NNNLO} is seen to deviate from {\NNLO}. We note that the $\cd{3}{3}$ phase shifts are dominated by iterations of OPE, as demonstrated by the deltaless $\cd{3}{3}$ in Fig.~\ref{fig:d_deltaless} and Fig.~9 in Ref.~\cite{Kaplan:2019znu}. In addition, the difference is exaggerated somewhat by the relatively small scale in the $\cd{3}{3}$ plot. 

Compared with the deltaless TPEs, the delta isobar provides much needed central forces, as exhibited in spin-singlet channels, e.g., $\cd{1}{2}$. At {\NNNLO}, the newly promoted isoscalar central counterterm $V_C^{(q^4)}$ gives us a handle to reduce the attraction of $V_{C-\Delta}^\text{TPE}$ towards higher momenta. A more optimum value of $D_{C-\Delta}$ could be obtained by a combined fit to various $D$-wave channels.    

\begin{figure}[htb]
    \centering
    \includegraphics[scale=0.49]{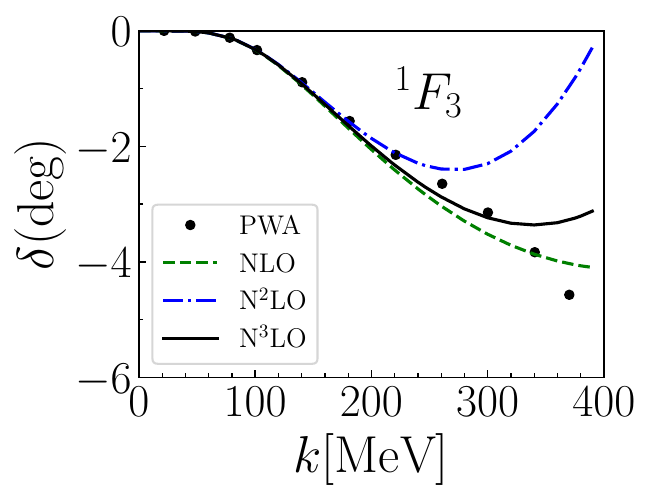}
    \includegraphics[scale=0.49]{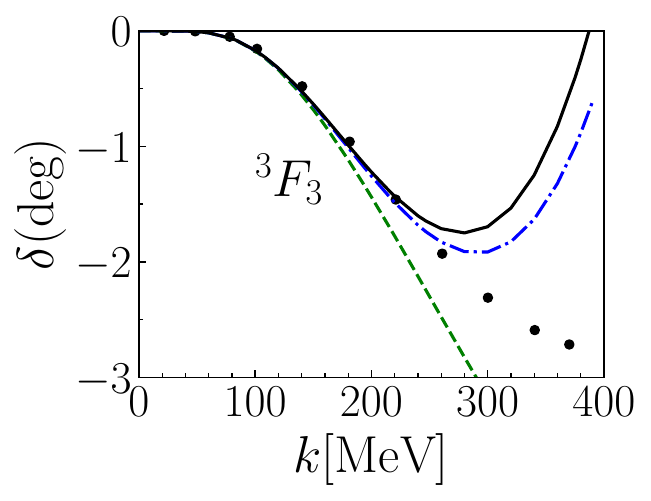}
    \includegraphics[scale=0.49]{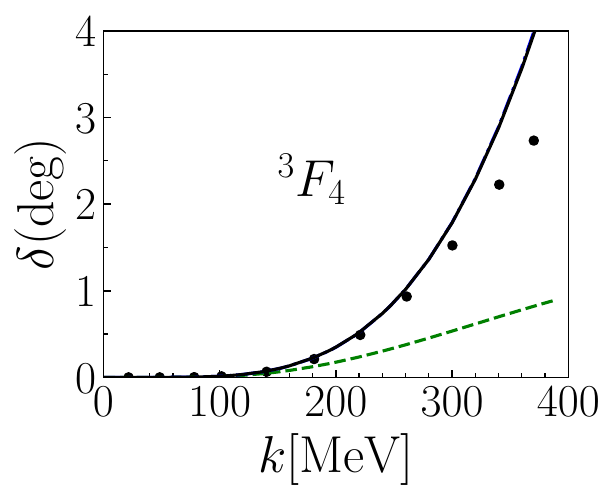}
    \caption{$NN$ phase shifts as functions of CM momentum in various $F$ waves with the deltaful TPEs.\label{fig:f_newC4}}
\end{figure}

\begin{figure}[htb]
    \centering
    \includegraphics[scale=0.49]{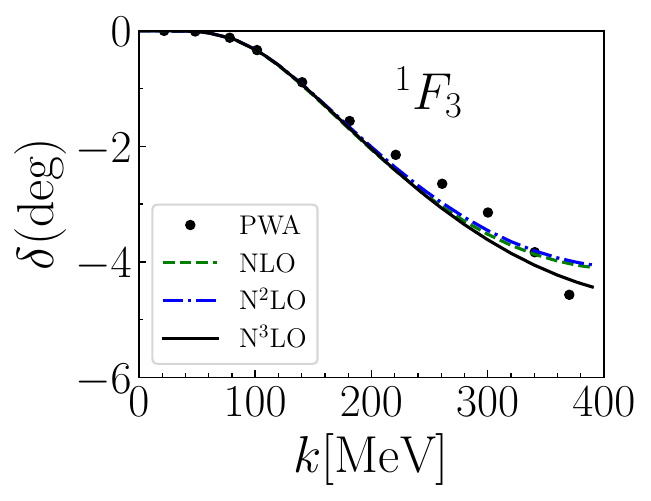}
    \includegraphics[scale=0.49]{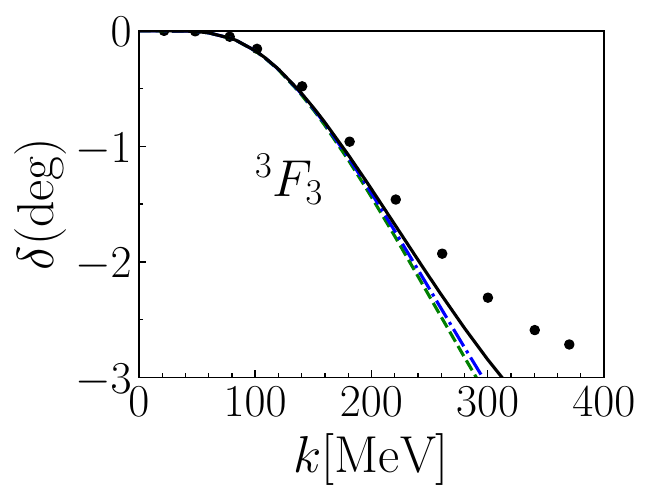}
    \includegraphics[scale=0.49]{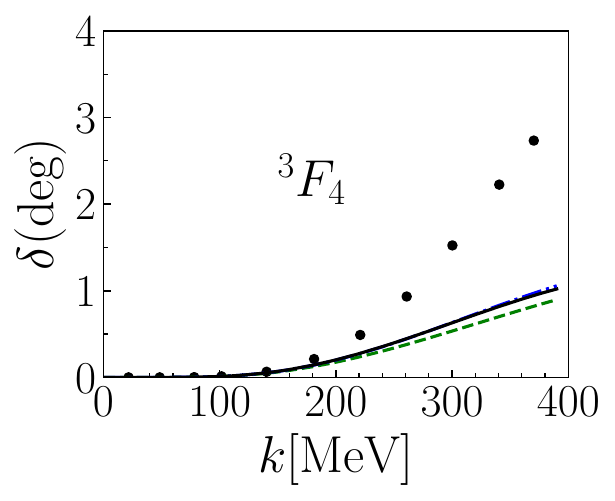}
    \caption{$NN$ phase shifts as function of CM momentum in the $F$ waves with the \emph{deltaless} TPEs.}
    \label{fig:f_deltaless}
\end{figure}

Because the counterterm $V_C^{(q^4)}$ does not act in the $F$ waves, there is no need to tune its value there. For the same reason to have excluded $\cd{3}{1}$, we do not include $\cf{3}{2}$, as OPE could become strong enough to be nonperturbative in $\cpf$. In Figs.~\ref{fig:f_newC4} and ~\ref{fig:f_deltaless}, one finds that the significant attraction provided by the deltaful central TPE remains the most prominent feature. In addition, we notice that the deltaless TPEs have little effects in the $F$ waves by agreeing with the OPE phase shifts, while the deltaful TPEs turn away from the OPE curve (NLO) considerably.

\section{Summary and Discussions\label{sec:disc}}

Although the delta isobar has been long speculated to play a role in chiral nuclear forces due to its proximity in mass to the nucleon, incorporating $\Delta$ into the hierarchy of chiral forces is plagued by a few issues. The dimensionally regularized deltaful TPE potentials, first derived in Ref.~\cite{Kaiser_1998}, have an isoscalar central component that appears too strong for a force considered two orders smaller than OPE. Later a different regularization scheme--- the spectral function regularization--- was advocated in Ref.~\cite{Krebs:2007rh}, and a particular choice of the cutoff value $\tilde{\Lambda} = 700$ MeV was made in order to agree with the $NN$ phase shifts. At the beginning, we investigated the reasons behind the significant disparity between $\tilde{\Lambda}=700$MeV and $\infty$ in SFR and why DR potentials are so strongly attractive compared with SFR. 

Assisted by the analytic expressions of the SFR potentials given in Ref.~\cite{Krebs:2007rh}, we were able to analyze the counterterms subsumed in the deltaful TPEs and found that the $q^4$ counterterm associated with the isoscalar central force $V_{C-\Delta}^\text{TPE}$ has a numerical factor surprisingly large, as shown in Eq.~\eqref{eqn:VCq4}. As a counterterm, $V_C^{(q^4)}$ is smaller than the long-range part of $V_{C-\Delta}^\text{TPE}$ by $\tilde{\Lambda}^{-1}$; therefore, we realized that $V_{C-\Delta}^\text{TPE}$ is more important than in Weinberg's scheme. That is, even though it is a one-loop diagram, $V_{C-\Delta}^\text{TPE}$ is suppressed by $Q^2/(4\pi f_\pi^2)$ relative to OPE, instead of $Q^2/(4\pi f_\pi)^2$. 

\begin{figure}[htb]
    \centering
    \includegraphics[scale=0.8]{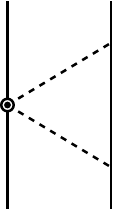}
    \caption{
    The seagull diagram for the TPE potential with a $\pi \pi NN$ vertex. For explanation of other symbols, see Fig.~\ref{fig:OneDeltaBox}.}
    \label{fig:Triangle}
\end{figure}

Examining the counterterms helped identify numerically the dominant long-range forces, but the origin of this enhancement remains unclear. While we did not seek to answer that question in the current paper, we surmise that this is related to the enhancement of the triangle diagram noticed by Refs.~\cite{Liebig:2010ki, Baru:2012iv}, shown in Fig.~\ref{fig:Triangle}. If the $\pi \pi NN$ vertex is momentum dependent, a similar enhancement arises: in the denominator, there is a factor of $\pi$ instead of $\pi^2$. (We note that the $\nu = 0$ $\pi \pi NN$ vertex, i.e., the famed Weinberg-Tomozawa term, is energy dependent.) The connection between the triangle diagram and Fig.~\ref{fig:OneDeltaBox} (b) is that when the delta pole is picked up in Fig.~\ref{fig:OneDeltaBox} (b), the resultant three-momentum integral resembles that of the nucleon-pole contribution of the triangle diagram.

The promotion of $V_{C-\Delta}^\text{TPE}$ to {\NNLO} in peripheral waves also means that one has to adjust the counting of $V_{C}^{(q^4)}$ by assigning it to {\NNNLO}, one order down from NDA. This modification of power counting affords us a handle to moderate the attraction of $V_{C-\Delta}^\text{TPE}$ before going up to {\NNNNLO}. 

The phase shifts up to {\NNNLO} showed in general a stronger attraction than in the absence of the delta isobar. While this is welcome in some channels ($\cd{1}{2}$, $\cf{3}{4}$), the attraction seems excessive in others ($\cd{3}{2}$, $\cf{3}{3}$). Going forward, there are several improvements we plan to implement. The value of the $N \Delta$ coupling $h_A$ and the nucleon-delta mass splitting $\Delta$ may be adjusted in conjunction with the analysis of $\pi N$ scattering. To be consistent with the promotion of $V_{C-\Delta}^\text{TPE}$ on grounds of factors involving $\pi$, the triangle diagram with $\nu = 1$ vertexes shown in Fig.~\ref{fig:Triangle} is to be promoted by one order as well.

\acknowledgments
This work was supported by the National Natural Science Foundation of China (NSFC) under Grant Nos. 12275185, 12335002 (BL), and 12347154 (RP).

\bibliography{refs.bib}

\end{document}